\documentclass{he_symp}
\usepackage{psfig,graphicx,epsfig}
\usepackage{color}
\usepackage{amsmath,amssymb,epic,eepic,array}
\begin{document}
\renewcommand{\FirstPageOfPaper }{ 136}\renewcommand{\LastPageOfPaper }{ 141}
\newcommand{\eqb}{\begin{eqnarray}}
\newcommand{\eqe}{\end{eqnarray}}
\newcommand{\diff}{{\rm d}}
\newcommand{\Lspin}{L_{\rm spin}}
\newcommand{\obsdir}{\vec{\hat{n}}}
\newcommand{\rsheet}{r_{\rm s}}
\newcommand{\rlight}{r_{\rm L}}
\newcommand{\nlight}{n_{\rm L}}
\newcommand{\blight}{B_{\rm L}}
\newcommand{\rstart}{r_0}
\newcommand{\degree}{{\rm o}}
\newcommand{\sigmaT}{\sigma_{\rm T}}
\newcommand{\Pdot}{\dot{P}}
\newcommand{\gtear}{\gamma_{\rm t}}
\newcommand{\gsynch}{\gamma_{\rm s}}
\newcommand{\ldebye}{\lambda_{\rm D}}
\newcommand{\rcrit}{r_{\rm cr}}
%
%

\title{A wind model for high energy pulses}
\author{J. G. Kirk\inst{1} 
 \and O. Skj\ae raasen\inst{1}
 \and Y. A. Gallant\inst{2}
}
\institute{Max-Planck-Institut f\"ur Kernphysik,
Postfach 10 39 80, 69029 Heidelberg, Germany
\and
Service d'Astrophysique, CEA-Saclay, F--91191 Gif-sur-Yvette, France}
\maketitle
\abstract{A solution to the $\sigma$ problem --- 
that of finding a mechanism capable of
converting Poynting energy flux to particle-borne energy flux in a pulsar 
wind --- was proposed several years ago by Coroniti and Michel who
considered a particular prescription for 
magnetic reconnection in a striped wind. This 
prescription was later shown to be ineffective. In this paper,
we discuss the basic microphysics of the reconnection process 
and conclude that a more rapid prescription is permissible. Assuming 
dissipation to 
set in at some distance outside the light-cylinder, we compute
the resulting radiation signature and find 
that the synchrotron emission of heated
particles appears periodic, 
in general showing both a pulse and an
interpulse. The predicted spacing of these agrees 
well with observation in the case of the Crab and Vela pulsars.
Using parameters appropriate for 
the Crab pulsar --- magnetization parameter at the light cylinder 
$\sigma_{\rm L}=6\times10^4$, Lorentz factor
$\Gamma=250$ --- reasonable agreement is found with the 
observed total pulsed luminosity. 
This suggest that the high-energy pulses
from young pulsars originate not in the co-rotating magnetosphere 
within the light cylinder (as in all other models) but from the
radially directed wind well outside it. 
}
\section{Introduction}
Recently, the non-axisymmetric \lq\lq striped pulsar wind\rq\rq\ investigated by
\cite{coroniti90} 
and 
\cite{michel94} 
has been reexamined (\cite{kirklyubarsky01}). 
A striped wind is a structure similar to a
Parker spiral, which is produced by the outward radial 
advection of magnetic field lines whose foot points 
are anchored on the surface of the rotating neutron star.
However, if the field lines originate from the poles of an 
obliquely rotating dipole, there exists 
a region around the equatorial plane where
the polarity of the field at a fixed radius reverses at
the rotation period of the star, producing stripes of alternating
magnetic field direction.
Magnetic energy is dissipated into particle energy if reconnection occurs
at the stripe boundaries. 
\cite{coroniti90} 
proposed that this kind of
dissipation proceeds at a rate sufficient to maintain the thickness of the
current sheet approximately equal to the gyro radius of the heated particles.
\cite{michel94} used essentially the same criterion, phrased in terms of the
velocity of the current carriers. These early papers concluded that 
observations of the Crab Nebula, which indicate conversion of Poynting flux
to particle-borne energy flux within $10^9\rlight$, 
(where $\rlight$ is the radius of the light cylinder)
could be explained by this
mechanism. However, \cite{lyubarskykirk01} point out that reconnection is
accompanied by acceleration of the wind. 
Both processes occur on the
same timescale when the Coroniti/Michel prescription is used for
reconnection. This is also the timescale on which the plasma
expands.
As a result, time dilation
in the accelerated flow has to be taken into account and the
dissipation rate is significantly reduced --- in the case of the Crab,
conversion of Poynting flux to particle-borne energy flux is not achieved until
roughly $10^{12}\rlight$, which is too slow to account for the
observations
(the so-called $\sigma$-problem; see \cite{melatos98} and
references therein).
  
In Sect.~\ref{sheet} we discuss the physics of the dissipation  
process and the prescriptions used to model it. We conclude that 
it may proceed much more rapidly than
suggested by Coroniti and Michel. 
If rapid reconnection in fact occurs, it will be accompanied
by radiation losses (Melatos 1998) that may give rise to an observable
signature. 
In this paper, and in an accompanying letter (Kirk et al.~2002)
we adopt this hypothesis.
We assume
that dissipation is 
triggered at a surface $r=\rstart$
well outside the light cylinder 
in the supersonic MHD wind of a pulsar
and takes place rapidly,  
so that the plasma does not have time to expand 
significantly.
Using two simple models of the electron distribution we 
show (i) that for parameters thought typical of the winds from young
pulsars
(e.g., the Crab) the predicted emission is pulsed, with, in general, a pulse
and interpulse that are not symmetrically spaced in
phase and (ii) that the predicted synchrotron 
luminosity in the pulses agrees to
order of magnitude with that observed from the high energy emission of
the Crab pulsar. 

Our calculations suggest that the 
high energy pulses emitted by young pulsars 
originate as synchrotron emission in the pulsar wind, far outside the light
cylinder. 
This is in marked contrast with conventional models of 
high-energy pulse production
at an inner or outer gap --- for a review see
\cite{harding01} --- and also with magnetospheric synchrotron based models
\cite{machabelietal00, crusiuswaetzeletal01}. The 
current sheet just beyond the region of closed field lines in the magnetosphere 
has also
been proposed as the source of high energy radiation from the 
Crab pulsar (Lyubarsky1996). But, in common with the 
others, this model relies for pulse production 
on the co-rotation with the star of a beam of radiation
emitted from within the light cylinder.

\section{Dissipation in the current sheet}
\label{sheet}
An equilibrium solution of the Maxwell-Vlasov equations describing 
a magnetic reversal across a current sheet was found many years ago (Harris 1962)
and was subsequently generalized to apply also to the case of 
relativistic particle temperatures and drift speeds (Hoh 1968).
If, in the special case in which the plasma is composed
solely of positrons and electrons
the magnetic field strength far from the sheet is $B'$, 
the central 
density of particles of each charge is $N'$ and the sheet width is $a$, 
(all measured in the frame of reference in which the sheet is stationary
and the electrons and positrons have equal and opposite drift speeds) then 
the temperature $T$, which is independent of position in the sheet, 
is given by

\eqb
T&=&\frac {{B'}^2} {16\pi N'}
\label{pbalance}
\eqe\\

\noindent
The Debye radius is

\eqb
\ldebye&=&\left(\frac {T} {4\pi N' e^2}\right)^{1/2}
\label{ldeb}
\eqe\\

\noindent
and is, to within a factor of the order of unity, equal to the gyro radius of
a particle of energy $T$ in the magnetic field $B'$.
The speed $\pm c\beta$ with which each component drifts is also independent 
of position in the sheet 
and is
obtained from 

\eqb
\beta&=&\frac {\ldebye} {a}
\label{speed}
\eqe\\

\noindent
The magnetic field reverses across the sheet and the density 
goes to zero far from the sheet --- simple explicit expressions 
for these quantities as functions of position are available.

This solution is well-known to be unstable 
to the excitation of a variety of waves --- see, for example, 
\cite{daughton99}. The standard scenario is that the 
current sheet first filaments because of the growth of the tearing mode,
producing magnetic islands separated by neutral lines. These then lead to
magnetic reconnection and its associated dissipation (Melrose 1986).
This scenario has largely been developed in the context of the geomagnetic 
tail of the Earth. Particle-in-cell simulations appear to 
confirm its validity ([Zhu \& Winglee 1996; Pritchett et al.~1996), 
although the 3-dimensional picture is more complicated (B\"uchner \& Kuska 1999;
Hesse et al.~2001).

The key question in the application to pulsar winds is 
that of how quickly 
and, therefore, whereabouts in the wind 
the dissipation occurs. The linear growth rate $\gtear$ 
of the tearing instability has been calculated in the relativistic case 
by \cite{zelenyikrasnoselskikh79}. For $T\gg mc^2$ they find 
for the most unstable mode,
which has a wavelength approximately equal to the sheet thickness,

\eqb
\gtear&=&\beta^{3/2}c/a = \ldebye^{3/2}c/a^{5/2}
\eqe\\

\noindent
The magnetic field far from the sheet and the total number of 
particles per unit area of sheet ($\approx aN'$) suffice to determine
the drift speed.
A single additional parameter is then needed 
to fix the sheet properties. This can
be chosen, for example, as the thickness $a$ or the temperature $T$;
the thinner (cooler) the sheet, 
the faster the tearing mode grows.

In his model of the high-energy emission from pulsars, \cite{lyubarsky96}
argues that the rate of dissipation is controlled by the 
timescale associated with the growth of the tearing mode. Specifically, 
he assumes that magnetic flux is swept into the sheet and annihilated at 
a speed given by $a\gtear$. This scenario is plausible, although it is not 
certain that the field dissipation, which presumably occurs at null points
created during the field filamentation, proceeds on the timescale of 
the tearing mode growth. If it operates, it results in
strong radiation from the innermost parts of the sheet. The geometry of 
the flow in this region is, however, uncertain.

Another, simpler, prescription for the speed of dissipation was 
(implicitly) adopted by both \cite{coroniti90} and \cite{michel94}
and, later, also by \cite{lyubarskykirk01}. In terms of the magnetic field and
the number of particles per unit area in the sheet, the drift speed 
can be written

\eqb
\beta&=&\frac {B'} {8\pi a N' e}  \label{drift}
\eqe\\

If the wind expands radially without any dissipation ($a=\,$constant), then 
its speed remains constant so that $B'\propto r^{-1}$ and $N'\propto
r^{-2}$. Equation~(\ref{drift}) 
then implies that the drift speed $\beta$ grows linearly with radius.
At a radius where, according to this equation, $\beta>1$, no equilibrium of 
the Harris type can exist; the 
Coroniti/Michel prescription
consists in assuming that the sheet thickness grows at a rate just sufficient
to avoid this happening. Dissipation is thus assumed to maintain 
the current sheet very close to an equilibrium in which $\beta\approx1$.
This enables reconnection to be incorporated
straightforwardly into a set of MHD equations describing the pulsar
wind. Provided the sheet does not radiate significantly, conservation of 
energy, momentum and particle number, together with Eq.~(\ref{speed}) in which
$\beta$ is set to unity,
determine the 
dynamics of the wind. However, in this case dissipation proceeds on
a time scale corresponding to the length $r$, which is the same scale on which 
adiabatic expansion of the sheet operates. A careful analysis of the dynamics 
shows that the wind accelerates and, as a result, 
the dissipation becomes rather ineffective (Lyubarsky \& Kirk 2001).

On the other hand, an instability in the current sheet can communicate itself
to the rest of the pulsar wind at the speed of the fast magnetosonic wave,
which, for a cold plasma is roughly equal to the Alfv\'en speed.
Recent investigations of 
reconnection in a relativistic pair plasma (Zenitani \& Hoshino 2001)  
indeed indicate that the 
relevant timescale is related to the crossing time of 
a magnetosonic wave. 
In the case of a transsonic pulsar wind, a magnetosonic wavefront could appear almost
stationary in the frame of the observer, so that dissipation could be 
completed within a few pulsar periods. This is the prescription we adopt in
this paper. Energy is assumed to be released in the current sheet at a fixed
radius, the position of which is taken as a
free parameter. After energy release, we assume the plasma cools slowly as the 
wind moves radially outwards.

\section{Geometry of the emission region}
\label{geometry}

Any radial, relativistic flow containing a periodic modulation of the
emissivity is likely to appear pulsed to an observer because of two 
effects.
Firstly, the strong
Doppler boosting of approaching parts of the flow means that only a small
cone of the flow is visible, propagating within an angle of roughly $1/\Gamma$
with respect to the line of sight, 
where $\Gamma=(1-\beta^2)^{-1/2}$ is the Lorentz factor and 
$c\beta$ the radial
speed. The
spread in arrival times at the observer of photons emitted at the same radius 
but at different parts within this cone is roughly $r/(c\Gamma^2)$. Secondly,
if emission is restricted to a range $\Delta r$ in radius, the fact that the
photons move only slightly faster than the flow when seen in the laboratory
frame means that the spread in arrival times is roughly $\Delta r/(c\Gamma^2)$.
For a wind modulated at the period $P=2\pi\rlight/c$ of a rotating star, 
the approximate 
condition for the observation of pulsed emission is therefore

\eqb
\Delta r,\ r&\la&\Gamma^2\rlight
\label{approximatepulse}
\eqe
\begin{figure}
\resizebox{\hsize}{!}
{\includegraphics[bb=74 278 705 507]{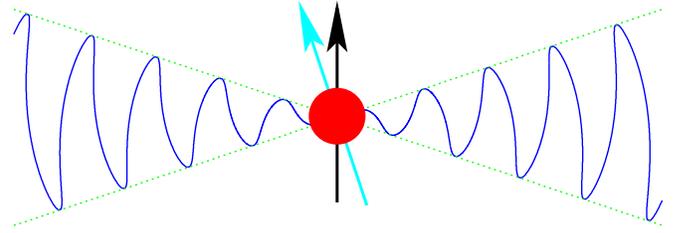}}
\caption{\protect\label{oblique}
A meridional section of the current sheet of an oblique
rotator
with a radial wind.
The thin lines limiting the extent of the sheet in latitude 
depict the plane of the magnetic equator when the magnetic axis
lies in the plane of the section.}
\end{figure}

\noindent
If this is fulfilled, the observed pulse shapes appear similar to the 
pattern of modulation of the flow. In particular, for a flow containing
the corrugated spiral current sheet of an oblique rotator
(Fig.~\ref{oblique}), there should in
general be two components (pulse and interpulse) which are symmetrically placed
if the pulsar is viewed from the equatorial plane, but are asymmetrical when
viewed from higher latitudes. 
Denoting by $\alpha$ the angle between the
rotation axis and the magnetic axis, and by $\zeta$ the angle between the
rotation axis and the viewing direction, these pulses appear only for 
$90^\degree -\alpha < \zeta < 90^\degree + \alpha$. 

To improve on the estimate (\ref{approximatepulse}) and 
calculate the expected pulse shapes we adopt 
the following 
model of the volume emissivity $\epsilon$ of the wind:

\begin{itemize}

\item[(i)] The emission region is located on the current sheet in the 
super-magnetosonic MHD wind
of a rotating neutron star with an oblique, split-monopole magnetic
field. 

\item[(ii)] The dependence on frequency $\nu$ of the volume emissivity is a power-law: 
$\epsilon\propto\nu^{-a}$.

\item[(iii)] The emission switches on abruptly as plasma in the sheet crosses the surface
$r=\rstart$. Thereafter the volume emissivity decays as 
a power law in radius: $\epsilon\propto r^{-2-q}$.
\end{itemize}

Outside the light cylinder, assumption (i) 
should be a good approximation,
independent of the actual configuration of the magnetic
field close to the pulsar surface.  
The assumption that emission occurs only at the
exact position of the current sheet is an idealization --- in reality, the entire flow may be
modulated, making the observed pulse more complex. 
In this case, our computations provide the Greens
function with which the pattern is to be folded to find 
the observed pulse shape. 
In the ultra-relativistic limit, a simple
wind solution is available (Bogovalov 1999) in which the 
velocity is radial and constant and the magnetic field is almost toroidal.
The thin current sheet in such a wind is described by the equation

\eqb
r&=&\rsheet(\theta,\phi,t)
\nonumber\\
&=&\beta\rlight\left[
\pm\arccos\left(-\cot\alpha\cot\theta\right)+c t/\rlight -\phi +2n\pi
\right]
\label{sheeteq}
\eqe\\

\noindent
(\cite{kirklyubarsky01}), where ($r$,$\theta$,$\phi$) are polar coordinates,
$t$ is the observer-frame time and $n$ is an integer.
 
Assumption (ii) is an idealization that ensures the
pulse shape is independent of frequency. In reality, the electron distribution
may exhibit breaks and cut-offs, which will be translated into different pulse
shapes at different frequencies. Our assumption, however, ensures that all
parts of the flow can contribute to the emission at a given frequency,
making it harder to obtain sharply defined pulses. 

\begin{figure}
\resizebox{\hsize}{!}
{\includegraphics[bb=40 159 566 689]{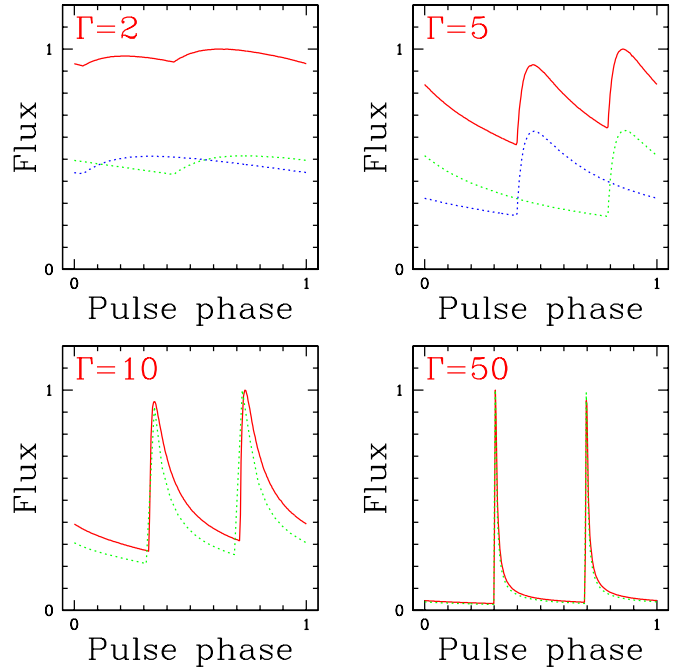}}
\caption{\protect\label{pulsefig}
Pulse shapes obtained by a numerical integration of Eq.~(\protect\ref{fluxeq}).
Four values of the Lorentz factor ($\Gamma$) are shown, 
the remaining parameters are: 
$\alpha=60^\degree$, $\zeta=60^\degree$, $\rstart=30\rlight$, $a=0$
and $q=3$ (see text). The total emission is shown as a solid line.
For $\Gamma=2$ and $4$, the contributions of the current sheets of different polarity
are shown as dotted lines.
For $\Gamma=10$ and $50$, the dotted line shows 
the effect of a more rapid fading of the
emissivity to larger radius ($q=6$).}
\end{figure}

The dependence of the observed flux
$F$ on $t$, $\alpha$ and the direction of observation $\obsdir$ is:

\eqb
F&\propto&
\int_{-\infty}^{+\infty}\diff t'\int_{\rstart}^\infty \diff r
\int_{90^\degree-\alpha}^{90^\degree+\alpha} \diff\theta\int_0^{2\pi}\diff\phi
\nonumber\\
&&r^{-q}D^{2+a}\delta\left[r-\rsheet(\theta,\phi,t')\right]
\delta\left(t'-t+\obsdir\cdot\vec{r}/c\right)
\label{fluxeq}
\eqe\\

\noindent
where $D=1/\left(1-\beta\vec{\hat{r}}\cdot\obsdir\right)$ is the 
conventionally defined Doppler 
factor, apart from a (constant) factor of $\Gamma$.
Choosing $a=0$ and $q=3$, which is appropriate
for the synchrotron radiation of a fixed number of particles that undergo
predominantly adiabatic losses in the expanding plasma,
we present sample pulses for various Lorentz factors in
Fig.~\ref{pulsefig}. For the higher Lorentz factors this figure also
displays the effect of a faster fall-off of the emissivity towards
higher radii ($q=6$), mimicking the effect of an accelerating wind. The
asymmetry in time of the pulses is produced by our assumption of a
sudden switching on 
of the emission and is reduced by a more rapid switching off.

\noindent
The parameters of this figure have been chosen to correspond to the Crab
Pulsar. 
The radio emission of this object is complex, but 
is known to display \lq
core emission\rq\ in the radio precursor
(Rankin (1990), which strongly suggests that the magnetic
axis passes within a few degrees of the line of sight. This indicates that the 
inclination angle $\alpha$ of the magnetic axis to the rotation axis
roughly equals the angle $\zeta$ 
between 
the rotation axis and the line of sight. The latter  
is determined by X-ray and
optical observations of the \lq torus\rq\ 
(Aschenbach \& Brinkmann 1975; Hester et al.~1995) to be $60^\degree$. 
In Fig.~\ref{pulsefig}
the emission has been assumed to switch-on at $\rstart=30\rlight$ 
and pulses appear for $\Gamma>5$, in rough agreement with 
the estimate (\ref{approximatepulse}). At higher Lorentz factors, the 
pulses sharpen. From Eq.~(\ref{sheeteq}) it can be seen that the
emission splits naturally into two contributions, corresponding to 
different signs of the term $\arccos(-\cot\alpha\cot\theta)$. In
the first two plots of Fig.~\ref{pulsefig} these contributions are plotted
separately as dotted lines. 
Although the detailed pulse shapes will depend on the
structure of the reconnecting sheets in the Crab wind, as well as the 
details of the switching on mechanism and the amount of acceleration
of the wind, the separation 
of the two peaks is not sensitive to these unknowns, 
being determined by the location of the
sheets. This separation in phase equals 
$\arccos(\cot\alpha\cot\zeta)/180^\degree$,
corresponding to $0.39$ for the Crab Pulsar, 
in good agreement with observation (e.g., \cite{kanbach98}).

In the case of the Vela pulsar, conflicting interpretations of the 
polarization sweep and the
X-ray morphology have yielded $\alpha\approx71^\degree$,
$\zeta\approx65^\degree$ in one case (Radhakrishnan \& Deshpande 2001) 
and
$\alpha\approx\zeta=53^\degree$ 
in the other
(Helfand et al.~2001). These imply separations of the
gamma-ray (sub-)pulses of $0.45$ and $0.31$ of a period respectively, 
so that the observed 
spacing of $0.4$ marginally favors the former interpretation.

\section{Luminosity estimates}
\label{luminosity}
Electrons heated in the current sheets will enter the
surrounding magnetic field and emit synchrotron radiation. A rough estimate
of the power radiated 
can be found by assuming a monoenergetic electron population 
of Lorentz factor $\gamma$ ($\gg1$) and
requiring pressure
equilibrium between the hot plasma and the surrounding magnetic field. 
Taking the single particle synchrotron emissivity to be a delta
function in frequency, the luminosity at frequency $\nu$ into 
a solid angle $\Omega$ is given by

\eqb
\frac {\diff L} {\diff\nu\diff\Omega}&\approx&\int_{\rstart}^\infty\diff r\,
r^2 \Delta\Gamma n'_{\rm h}P_{\rm s.p.}\delta\left(\nu-\nu_0\right)
\label{logos}
\eqe\\

\noindent
(Skj\ae raasen \& Kirk 2001), where $\Delta$ ($\le1$) is the fraction of a
wavelength of the wind pattern occupied by hot electrons of proper density
$n'_{\rm h}$, $P_{\rm s.p.}=(4/3)c\sigmaT\gamma^2({B'}^2/8\pi)$
is the synchrotron power emitted by a 
single electron, and $\nu_0$ is the characteristic frequency of emission, 
measured in the observer frame ($\sigmaT$ is the Thomson cross section). 
The condition of pressure equilibrium gives
$n'_{\rm h}\gamma mc^2/3= {B'}^2/8\pi$, where $B'$ is the magnetic field in
the co-moving frame and $m$ the electron mass. 
Well outside the light cylinder, but 
before energy release begins; i.e., at $\rlight<r<\rstart$, 
the density and magnetic field scale as $r^{-2}$ and $r^{-1}$ respectively, 
and can be characterizes by the values $\nlight$ and $\blight$ 
obtained by extrapolating 
back to $r=\rlight$. Provided the flow does not accelerate significantly 
whilst emitting synchrotron radiation, these same scalings,
which imply $\gamma$ independent of $r$, can be used 
in the integration in Eq.~(\ref{logos}). 
This is convenient for a rough estimate of the luminosity, but is sensitive to
the assumption that the wind does not accelerate upon reconnection.

If we identify the ratio of Poynting flux to kinetic energy (at the
light cylinder)
$\sigma_{\rm L}=\blight^2/(4\pi\Gamma \nlight mc^2)$, 
define an average fraction of hot particles
$\bar{\Delta}=\rstart\int_{\rstart}^\infty\diff r\Delta
/r^2\la1$, and use the standard relation
$\blight=2.9\times10^8\Pdot^{1/2}P^{-5/2}\,$G, the luminosity becomes

\eqb
\frac {\diff\hat L} {\diff \Omega}
&=&
3.9 \times10^7\left(\frac {\sigma
\bar{\Delta}\rlight\Pdot} {\Gamma^3\rstart P^4}\right)
\ (P\ {\rm in\ secs})
\label{finalflux}
\eqe

\noindent
where $\diff\hat L/\diff \Omega$
is the synchrotron luminosity per steradian divided 
by the total wind luminosity per steradian.
Similarly, the characteristic frequency is

\eqb
h\nu_0(r)&=&
2.6\times10^{-6}\left(\frac {\rlight} {r}\right)\sigma^2\Pdot^{1/2}P^{-5/2}
\ ({\rm MeV}).
\label{freq1}
\eqe

\noindent
The spectrum produced by this simple model with
monoenergetic electrons extends to a maximum frequency
given by Eq.~(\ref{freq1}) with $r=\rstart$ and 
is flat:
$\diff L/\diff\nu\diff\Omega\propto \nu^0$.  However, this aspect is 
sensitive to the way in which the plasma expands and accelerates after 
reconnection.
If, in a more realistic 
scenario, the electron density is a power law 
$N(\gamma)\propto\gamma^{-p}$, with $p>2$, then most of the energy resides in 
the particles with lowest $\gamma$ and the above estimate (\ref{freq1}) 
refers to the photons emitted by these particles. The 
emission at frequencies $\nu<\nu_0(\rstart)$ 
is dominated by the same low energy particles
radiating at larger radius
in a weaker magnetic field, and is determined by the expansion and 
acceleration of the hot plasma. In the absence of acceleration, the flat
spectrum derived above
prevails.
However, the intrinsic spectrum, which is a power law 
$\diff L/\diff\nu\diff\Omega\propto \nu^{-a}$,
with $a>0.5$, is 
revealed at frequencies $\nu>\nu_0(\rstart)$, where particles
radiating at $\rstart$ dominate.
In the case of the Crab pulsar, 
the optical to X-ray emission is consistent with a flat spectrum
(Shearer \& Golden 2001), which steepens to $a>1$ 
at a photon energy of 
approximately $1\,$MeV (Kanbach 1998; Kuiper et al.~2001).
Equation~(\ref{freq1})
therefore
implies
$\sigma=1.1\times10^4\sqrt{\rstart/\rlight}$. Observations suggest
$\diff\hat L/\diff\Omega\approx 10^{-3}$ for this object, which,
from Eq.~(\ref{finalflux}), is consistent with a mildly supersonic flow,
$\Gamma\approx 250$, that starts to radiate at $\rstart=30\rlight$. 
This is consistent with the lower limit of $5\rlight$ found by 
(Bogovalov \& Aharonian 2000) for the radius at which the conversion of
Poynting flux to particle-born flux may occur.

\section{Discussion}
\label{discussion}
Current models of the high energy emission from rotation powered
pulsars fall loosely into two groups: \lq\lq polar cap\rq\rq\ and 
\lq\lq outer gap\rq\rq\ models (for a recent review see 
\cite{harding01}). Each of these places the emission region in the
corotating magnetosphere of the star. The predicted pulse shapes 
depend sensitively on the uncertain geometry of the magnetic field in
this region, which is usually assumed dipolar. 
In contrast, in the model
presented above, pulses are emitted well outside the co-rotation
region. Provided the pulsar drives a supersonic, MHD wind, the
field geometry approaches a simple asymptotic solution
(Bogovalov 1999), which determines the basic properties of the
pulses. 

In a series of papers,
\cite{pacini71,pacinisalvati83,pacinisalvati87} 
suggested that particles close to the light cylinder emit 
synchrotron radiation and are responsible for the optical pulses of the
Crab and other pulsars --- a theory which appears to be in reasonable
agreement with observation (Shearer \& Golden 2001).
The model presented here is similar in that the radiation
mechanism is synchrotron emission and the scaling of Eq.~(\ref{finalflux}) 
is close to that originally given by \cite{pacini71}. It goes further,
however, by specifying for the emission region
a precise geometrical structure and location.

It is known that the dilution of the plasma 
in the wind must lead to non-ideal MHD behavior  (Usov 1975, Michel 1982)
and so can be responsible for triggering the emission. 
The position at which this occurs depends on the 
initial concentration of charges in the current sheet and is taken as
a free parameter. The major uncertainty in the model is the speed with
which dissipation proceeds. Our computations implicitly assume rapid
dissipation (over a scale small compared to the radius). In this
case, it is possible to make a rough estimate of the synchrotron
luminosity which is in agreement with observations. A more detailed comparison 
with the wealth of relevant observations of optical, X-ray and
gamma-ray pulsars must await further modeling. 
It is also conceivable that
at least part of the radio emission could be produced in the wind region, 
as is suggested by the similarity of the pulse profiles 
at all frequencies seen in the Crab
pulsar. However, until a candidate coherent mechanism can be 
identified, this connection remains very speculative.
Nevertheless, the wind scenario is 
a viable alternative to current theories 
of gamma-ray, X-ray and
optical pulses from rotation-driven pulsars and has the advantage of making 
specific predictions of the pulse properties. 

\acknowledgement{This work is a collaboration of the TMR Network 
\lq\lq Astroplasmaphysics\rq\rq\ of the European Commission, 
contract FMRX-CT-98-0168. Y.G. is supported by a Marie Curie Fellowship 
from the European Community, contract no.\ HPMFCT-2000-00671 under the 
IHP programme. J.K. thanks the Wilhelm and Else Heraeus-Stiftung 
and the conference organisers for their invitation to attend and 
their hospitality at the meeting, and thanks the other participants for 
stimulating discussions.}


\clearpage

\end{document}